\documentclass[12pt]{article}

\usepackage{epsfig}

\usepackage{amssymb}
\usepackage{amsfonts}

\usepackage{color}
 
\def\be{\begin{equation}}
\def\ee{\end{equation}}
\def\ba{\begin{array}{c}}
\def\ea{\end{array}}

\newcommand{\bea}{\begin{eqnarray}}
\newcommand{\eea}{\end{eqnarray}}

\newcommand{\bbr}{\br\!\br}
\newcommand{\kkt}{\kt\!\kt}
\newcommand{\pbr}{\prec\!\!}
\newcommand{\pkt}{\!\succ\,}
\newcommand{\kt}{\rangle}
\newcommand{\br}{\langle}

\begin{document}


\vspace{.35cm}

\begin{center}

{\Large

Features, paradoxes and
amendments of perturbative
non-Hermitian quantum mechanics

 }

\vspace{10mm}

  {\bf Miloslav Znojil}$^{1,2,3}$

\end{center}

\vspace{10mm}

  $^{1}$
 {The Czech Academy of Sciences,
 Nuclear Physics Institute,
 Hlavn\'{\i} 130,
250 68 \v{R}e\v{z}, Czech Republic, {e-mail: znojil@ujf.cas.cz}}


 $^{2}$
 {Department of Physics, Faculty of
Science, University of Hradec Kr\'{a}lov\'{e}, Rokitansk\'{e}ho 62,
50003 Hradec Kr\'{a}lov\'{e},
 Czech Republic}

  $^{3}$
 {Institute of System Science, Durban University of Technology,
Durban, South Africa}

%
%
%
%
%
%



\section*{Abstract}

Quantum mechanics of unitary systems is considered in
quasi-Hermitian representation
and in the dynamical regime in which
one has to take into account
the ubiquitous presence of perturbations,
random or specific.
In the paper it is shown that multiple technical obstacles
encountered in such a context
can be
circumvented via just a mild amendment of the so called
Rayleigh-Schr\"{o}dinger perturbation-expansion approach.
In particular, the quasi-Hermitian
formalism characterized by
an enhancement of
flexibility
is shown to remain mathematically tractable
while,
on phenomenological side,
opening
several new model-building horizons.
It is emphasized that they include, i.a., the
study of generic random perturbations and/or of multiple specific
non-Hermitian toy models. In parallel, several paradoxes and open
questions are shown to survive.

\subsection*{Keywords}

unitary quantum mechanics;
non-Hermitian Schr\"{o}dinger picture;
generalized perturbation theory;
ambiguity of physical Hilbert space;

\newpage

\section{Introduction\label{introd}}

The exact or approximate solutions of time-independent
Schr\"{o}dinger equation
 \be
 H\,|\psi_n\kt = E_n\,|\psi_n\kt\,,\ \ \ \ |\psi_n\kt \in {\cal H}
 \,,\ \ \ \ n = 0,1,\ldots
 \label{sekvoj}
 \ee
play a key role in our understanding of the structure of quantum
bound states or resonances. Often, it is believed that
up to some truly exotic exceptions
the division line which separates the case of bound states
from the case of resonances
also separates Eq.~(\ref{sekvoj})
in which $H$ is Hermitian from Eq.~(\ref{sekvoj})
in which $H$ is non-Hermitian.
Incidentally, the latter belief has been shattered
after 1998 when
Bender with Boettcher \cite{BB}
revealed that the class of
the ``anomalous''
non-Hermitian Eqs.~(\ref{sekvoj}) yielding
bound states
can be larger than expected,
incorporating also certain
models in which the
Hamiltonians have the
form
of superposition
of the most common kinetic energy $\sim p^2$
with an equally standard (but complex)
local interaction potential.

In the latter models,
widely known as
``${\cal PT}-$symmetric'' \cite{BG,Carl,Carlbook,Christodoulides,Nimrodc},
the manifest non-Hermiticity of the Hamiltonian
 $$
 H \neq H^\dagger
 $$
has been found to
coexist with the
reality of the spectrum.
Thus, it was immediate to conclude that
the unitarity of the evolution
can be guaranteed
not only
in the conventional textbook spirit
(i.e.,
via the
self-adjointness
of the Hamiltonian \cite{Stone})
but also, under certain
additional technical conditions \cite{Geyer}, via the
Dieudonn\'{e}'s \cite{Dieudonne} quasi-Hermiticity
requirement
 \be
 H^\dagger\Theta=\Theta\,H\,.
 \label{zasdieu}
 \ee
One can then speak about quantum mechanics of unitary systems which
is slightly modified and reformulated in the so called
quasi-Hermitian representation.

In this framework
one of the most important features of the modification
may be seen in its innovative approach to
the concept of
perturbation which is found,
in this setting, counterintuitive. For three reasons. The first
one is that in this formalism (cf., e.g., its reviews in \cite{ali}
or \cite{book})
we are allowed to change the physical
Hilbert-space norm. Thus, in a preselected ``perturbed'' Hamiltonian
$H(\lambda)=H_0+\lambda\,H_1$ the size (and, hence, influence) of
the perturbation cannot always be kept under a reliable control.
Often, an enhanced sensitivity to perturbations is observed, for
this reason, in open quantum systems (a few more remarks on this
subtlety
will be added below).

The second
reason and paradox emerges
when we consider just
a closed quantum system in which the influence of $H_1\neq
H_1^\dagger$ is guaranteed to be small.
Still, the correct probabilistic
interpretation of the system remains ambiguous, mainly again due to the
non-uniqueness of the physical Hilbert-space inner-product
metric~$\Theta$ (again, a more detailed support of this
observation will be given below).

Thirdly, even if we decide to ignore the latter ambiguity
and even if we pick up just any one of the eligible metrics (which
would reduce the scope of the theory in a not quite predictable manner
of course), such a choice of the geometry of the
physical Hilbert space would
still vary with $\lambda$.
This is, probably, the
most challenging  problem which is to be also addressed
in our present paper.

Preliminarily we may notice and emphasize that
in the language of mathematics, the
problem may be formulated easily because the
underlying
auxiliary,
unitarity-of-evolution-guaranteeing operator $\Theta$
(if it exists \cite{Geyer}) can be perceived as
representing just an invertible and positive definite {\it ad hoc\,}
physical-Hilbert-space inner-product metric,
$\Theta=\Theta^\dagger >0$.
In the related reformulation of quantum theory all of the measurable
predictions only require, therefore, the evaluation of the following
metric-dependent matrix elements
 \be
 a_n =
 \br \psi_n|\Theta A |\psi_n\kt\,.
 \label{melasa}
 \ee
The
knowledge of the wave function
and of the operator $A$
representing an observable of interest
must be complemented here by the guarantee of
observability
$A^\dagger\Theta=\Theta\,A$ of course~\cite{Geyer,ali}.

One of the most influential sources of interest in
certain special
classes of non-Hermitian Hamiltonians with real spectra
lied in quantum field theory \cite{BM}
and, in particular,
in the role played there by
perturbation theory \cite{Wu,Turbiner,Kato}.
One of the most important subsets of
the underlying phenomenological Hamiltonians $H$
is formed, therefore, by the one-parametric families
 \be
 H=
 H(\lambda)=H(0) + \lambda\,V
 \neq H^\dagger\,
 \label{perham}
 \ee
where
$\lambda$ is a complex number and where
the component $V$ representing the perturbation should not be,
in some sense, too large \cite{Kato}.

Under this assumption,
a powerful tool of the construction
of the solutions of Schr\"{o}dinger Eq.~(\ref{sekvoj})~+~(\ref{perham})
lies in the use of power-series ansatzs
 \be
 E_n=E_n(\lambda)=E_n(0) + \lambda\,E_n^{(1)}+\lambda^2\,E_n^{(2)}+\ldots\,
 \label{20}
 \ee
 and
 \be
 |\psi_n\kt =|\psi_n(\lambda)\kt =|\psi_n(0)\kt + \lambda\,|\psi_n^{(1)}\kt
 +\lambda^2\,|\psi_n^{(2)}\kt+\ldots\,.
 \label{RSer}
  \ee
A serious obstacle emerges
when we turn attention to the unconventional
quasi-Hermitian models.
In the light of Eq.~(\ref{zasdieu})
the metric will become
manifestly $\lambda-$dependent in general,
$\Theta=\Theta(\lambda)$.
In contrast to the conventional
perturbation-expansion constructions,
it becomes necessary to complement
the standard pair (\ref{20}) and (\ref{RSer}) of the
Rayleigh-Schr\"{o}dinger perturbation-expanison
ansatzs by their new, operator-expansion
partner, say, of the
power-series form
 \be
 \Theta(\lambda)=\Theta(0) + \lambda\,\Theta^{(1)}
 +\lambda^2\,\Theta^{(2)}+\ldots\,.
 \label{2x}
 \ee
This means that non-Hermiticity (\ref{perham}) of  Hamiltonian
makes
a consequent implementation of the Rayleigh-Schr\"{o}dinger
perturbation-expansion
approach
to the quasi-Hermitian bound-state quantum physics
complicated.

The consequent theory requires
an explicit or implicit reference to
as many as five
separate
but mutually interrelated Hilbert spaces in general
(cf. paper \cite{response}).
The main theoretical benefit of such a
five-Hilbert-space reformulation of
quantum
mechanics
lies in an exhaustive classification
of admissible perturbations.
In this sense
our present paper
can be read as a
more pragmatically oriented follow up of
paper \cite{response}.

For introduction,
a few basic
features of the theory
may be found summarized in Appendices A -- D.
To this background, section \ref{aaaa} will add
two illustrative examples explaining not only
an overall motivation of using non-Hermitian Hamiltonians
but also the existence of the
deep mathematical differences between the
use of perturbation expansions in the closed
and open quantum systems.

In section \ref{violat} we will turn attention
to the physical consequences of these differences.
We will point out that
in the related literature
the necessity of an unambiguous
separation of the
closed-system  quantum physics
from the open-system quantum physics
is not always sufficiently
carefully observed.
This note
will be complemented by an outline of
the role of
random perturbations in realistic models.
A critique of a few
recent results will be given
in which the
depth of the difference between the
closed
and open systems has been underestimated.
This will be followed by a
clarification of
one of the related paradoxes
connected with the usage of the concept of
the so called pseudospectra \cite{Trefethen}.
For the description of the influence
of the random perturbations,
the usefulness of the concept of pseudospectra
remains
strictly restricted to the
studies of the open quantum systems.
In the quasi-Hermitian models
the transition from spectra to pseudospectra
cannot be recommended because it
does not lead to any enrichment of the information about
the dynamics of the underlying closed quantum systems.

In section \ref{whava} we will finally
return to the
quasi-Hermitian perturbation
theory. We will recall
the mathematical challenge
represented by the necessity of the construction of
an additional operator expansion
(\ref{2x}). In a climax of our paper we will
offer a new, alternative, upgraded
formulation of the quasi-Hermitian version
of the Rayleigh-Schr\"{o}dinger perturbation series
in which the latter necessity will be circumvented..

An extensive discussion and summary of our results will be
presented in the last two sections \ref{onintrod} and \ref{xonintrod}.
The essence of the innovation (and,
first of all,
of a significant simplification of the formalism)
will be shown to lie in an implementation of the
biorthogonal-basis ideas \cite{Brody} as
taken from their recent application
in a different, non-stationary
quantum dynamics context \cite{twoSE}.

\section{Merits of non-Hermitian Hamiltonians\label{aaaa}}

From a purely pragmatic point of view,
Schr\"{o}dinger Eq.~(\ref{sekvoj})
can be perceived
as a linear eigenvalue problem in which,
in the majority of applications, the
possible non-Hermiticity of the Hamiltonian would
make the construction of solutions
less stable and technically more difficult.
This
is a generic statement
which is well known \cite{book,Trefethen}.
The more the people seem
surprised when they encounter a
quantum system for which
the technically friendliest representation of
Hamiltonian happens to be non-Hermitian.

\subsection{Dyson-inspired simplifications of Schr\"{o}dinger equations}

A compact account of history of
the recent quick enhancement of interest
in the closed and stable quantum systems controlled by an
``anomalous'' Hamiltonian $H \neq H^\dagger$
can be found in \cite{book}.
The emergence of such a class of models
can look, at the first sight, surprising. Nevertheless, one of the oldest
demonstrations of the technical advantages of
using
a non-Hermitian $H$
already emerged
many years ago, viz., during the
Dyson's entirely pragmatic, well motivated and
purely numerical study
of a specific real-world many-body problem \cite{Dyson}.

An impact of the latter quantum-many-body
result remained, for a couple of years,
restricted just to nuclear physics \cite{Jensen}.
The idea only acquired a new life and broader response
after Bender with Milton \cite{BM}
revealed that the study of
non-Hermitian models
may be also of immediate relevance in quantum field theory.

In such a broadened methodical context
a particularly elementary
and fully non-numerical sample
$H^{(JM)}=p^2+{V}^{(JM)}(x)$ of such a
Hamiltonian
has been proposed,
in 2006, by Jones with Mateo
\cite{Mateo}.
Via
an
exactly solvable toy model these authors
demonstrated that in some cases,
given a conventional
self-adjoint
Hamiltonian
${\mathfrak{h}}^{(JM)}=p^2+{\mathfrak{v}}^{(JM)}(x)$,
one can profit from its replacement
by an isospectral alternative
$H^{(JM)}=p^2+{V}^{(JM)}(x)$
which is non-Hermitian.
Indeed, the former operator where the
potential contained
two components,
 \be
 {\mathfrak{v}}^{(JM)}(x)=
  - 2\,x
 +4\,x^4\,
 \label{bouha}
 \ee
could be perceived as
more complicated
than its avatar $H^{(JM)}$ with
 \be
 V^{(JM)}(x)= - \,x^4\,
 \label{hahab}
 \ee
containing
just the single interaction term. Moreover, the single-term
potential
(\ref{hahab})
is symmetric with respect to the
product of
parity and time-reversal, i.e.,
in comparison,
less complicated
than
its left-right-asymmetric two-term partner (\ref{bouha}).
One can really speak about a
simplification ${\mathfrak{h}}^{(JM)} \to H^{(JM)}$,
in principle at least.

Both of the latter
Hamiltonians
predict
the same real (i.e., measurable and stable)
bound-state spectrum which is discrete and
bounded from below.
The conventional one, viz., operator ${\mathfrak{h}}^{(JM)}$
is self-adjoint while its non-Hermitian avatar ${H}^{(JM)}$
is merely quasi-Hermitian (cf. definition (\ref{zasdieu})).
From an experimentalist's point of view
the isospectrality of the two alternative Hamiltonians
makes the two representations of the same
closed quantum system
indistinguishable.
For mathematicians,
the differences are also not too deep because
the
main source of difference, viz., the
inner-product metric needed in Eq.~(\ref{zasdieu})
has been found, in \cite{Mateo},
in an exact, closed and really elementary operator form
 \be
 \Theta^{(JM)}= \exp \left [
 p^3/48-2p
 \right ]\,.
 \label{[19b]}
 \ee
This makes the non-Hermiticity
of  ${H}^{(JM)}$
just a minor, easily surmounted complication.


\subsection{Analytic continuations and non-unitary open systems\label{baa}}

From the point of view of experimental
physics the truly exceptional
exact solvability of the Jones' and Mateo's interaction
(\ref{hahab}) is not so
impressive because
the coordinate $x$ is complex (see
its definition in \cite{Mateo}).
This makes the
standard probabilistic interpretation of the
``simplified'' system unclear
because the value of $x$
(tentatively treated as the position
of a particle)
ceases to be a measurable quantity.

A new physics has to be then sought
in a return to differential
Schr\"{o}dinger equations in which the coordinate
$x$ remains real.
In the light of the paradox (or rather of the danger) of the
non-observability of coordinates
one is forced to consider
the
asymptotically less anomalous
potentials in which
the line of coordinate $x$ could still be kept real.
One of such illustrative examples can be found
in our older
paper \cite{Rafa}
where we studied
the perturbation expansions (\ref{20})
of the energies generated by
the two-parametric
imaginary cubic oscillator Hamiltonian
 \be
 H^{(IC)}(f,g) =-\frac{d^2}{dx^2}+\frac{f^2}{4} x^2
 +{\rm i}g x^3\,.
 \label{1}
 \ee
Indeed, such a differential-operator model is still
non-Hermitian and
${\cal PT}-$symmetric, i.e., it is formally closely
analogous
to Eq.~(\ref{hahab}).
Moreover, knowing that
after identification $\lambda=g$, i.e.,
in the weak-coupling regime
the conventional small-anharmonicity
expansions would diverge,
we were able to transfer
the role of a small parameter
to the other coupling and choose
$\lambda=f$. As a consequence
we achieved a very good convergence
of our resulting perturbative
strong-coupling series (\ref{20})
for the energies.

Later, we found a complementary
inspiration in
paper \cite{BD}
in which Bender with Dunne set $f=1$ and $\lambda=g$ and studied
the alternative,
divergent but resummable weak-coupling expansion.
They were interested just in
the ground state energy,
and they managed to construct
the Rayleigh-Schr\"{o}dinger
perturbation series
 \be
 E^{(BD)}(\lambda) \sim \frac{1}{2}+\sum_{n=1}^\infty\,b_n\lambda^{2n}\,
 \label{2}
 \ee
up to very large orders by having
evaluated the necessary
integer-valued coefficients non-numerically,
 \be
 b_1=11\,,\
 b_2=-930\,,\
 b_3=158836\,,\
 b_4=-38501610\,,\
 \ldots\,.
 \label{3}
 \ee
At $n \gg 1$ they managed to
fit these coefficients by an amazing asymptotic formula
 \be
 b_n \sim (-1)^{n+1}
 \frac{60^{n+1/2}}{(2\pi)^{3/2}}\,\Gamma\left (n+\frac{1}{2}
 \right )
 \left [1+{\cal O}
 \left (\frac{1}{n}
 \right )
 \right]\,.
 \label{asyesi}
 \ee
Via an appropriate
resummation of the divergent series (\ref{2})
this enabled them to obtain,
at any not too large real coupling $\lambda$, a very good
(they even wrote ``excellent'')
agreement with the known and real
numerical value of the ground-state energy $E^{(BD)}(\lambda)$.

As a climax of the story, Bender with Dunne
proposed also a
phenomenologically meaningful
physical output
of their considerations. For this purpose
they re-interpreted their asymptotic estimate
(\ref{asyesi})
as a support of the possibility and consistency of
an analytic continuation of the function $E^{(BD)}(\lambda)$
to the (cut) complex plane of $\lambda$.
On these grounds they were able to
evaluate the
imaginary part of $E^{(BD)}({\rm i}\epsilon)$
and to
interpret the result as a prediction of a measurable decay width
of another quantum system described by
an analytically continued Hamiltonian
 \be
 H=p^2+x^2/4-\epsilon x^3
 \label{physi}
 \ee
(cf. Eq. Nr. 5 in \cite{BD}).
In other words, the initial
non-Hermitian operator (\ref{1})
has been reinterpreted, via analytic continuation,
as
a more or less standard
physical quantum Hamiltonian
supporting an unstable (but still
observable) ground state.


\subsection{Dyson maps and the modified concept of locality\label{moco}}

In the overall framework of
quasi-Hermitian quantum mechanics (QHQM) of closed systems
as formulated,
in Schr\"{o}dinger picture,
by Scholtz et al \cite{Geyer},
we paid attention,
in our recent paper~\cite{response},
to the consistent applicability of the theory
in the presence of perturbations.
We pointed out that even
in the
non-perturbative version of the
theory it made sense
to realize the description using three separate
Hilbert spaces
(cf. diagram Nr.~(10) in \cite{response}
or Eq.~(\ref{ufin}) in Appendix A below).

One of these spaces is denoted here by symbol ${\cal L}$.
By assumption, it is just a hypothetical
and, for practical purposes, inaccessible space.
Only the other two are relevant, both
hosting operator $H$
and differing just
by the
respective forms of inner products.
The first space (viz., ${\cal K}$) is just
auxiliary and unphysical. The second one
(denoted here as ${\cal H}$) is physical and, for this reason,
unitary equivalent to  ${\cal L}$, with
the equivalence mediated by a mapping
$\Omega$.

The latter (often called Dyson \cite{Dysonb})
mapping is related to the metric by formula
 \be
 \Theta=\Omega^\dagger\Omega \neq I\,.
 \label{fakju}
 \ee
In a way dating back to the Dyson's paper \cite{Dyson},
the key message as delivered by
our paper~\cite{response}
is that after one
makes the Hamiltonian $\lambda-$dependent
and after one
implements the perturbation-expansion philosophy,
one has to distinguish
between the ``physics'' (represented by
the perturbed $H(\lambda)$ at any $\lambda \neq 0$)
and ``mathematics'' (represented by
the exactly solvable $H(0)$).
In other words, both of the Hamiltonian-supporting
Hilbert spaces ${\cal K}$ and ${\cal H}$
become $\lambda-$dependent.

Incidentally,
at both $\lambda=0$ and $\lambda \neq 0$,
the knowledge of factorization (\ref{fakju})
would enable us to return also to the
above-mentioned toy-model interaction (\ref{hahab})
in which the ``false coordinate'' appeared to be complex,
$x \notin {\mathbb{R}}$.
Due to the exact solvability of the model
and due to the extreme simplicity of the related metric (\ref{[19b]}),
one could also introduce a closed-form Dyson-map operator
 \be
 \Omega^{(JM)}= \exp \left (
 p^3/96-p
 \right )\,
 \label{[19dm]}
 \ee
and {\em define\,}
a correct (i.e., by construction, quasi-Hermitian)
coordinate-representing operator $Q^{(JM)}$
acting in ${\cal K}$ and ${\cal H}$ using formula
 \be
 \mathfrak{q}^{(JM)}=\Omega^{(JM)}\,Q^{(JM)}\,\left (\Omega^{(JM)}
 \right )^{-1}=\mathfrak{q}^\dagger \,.
 \label{tarelax}
 \ee
This
is definition of a suitable (albeit a bit artificial)
observable tractable as a coordinate.
From the point of view of consistency of the theory
the choice of
self-adjoint $\mathfrak{q}^{(JM)}$ (or, directly, of
quasi-Hermitian $Q^{(JM)}$)
is more or less arbitrary.

Relation (\ref{tarelax}) itself
can be re-read
as the closest analogue
of connection between
the more common energy-operators
{\it alias\,} Hamiltonians
(cf. relation~(\ref{tarela}) in Appendix B).
Such a constraint can be complemented
by some additional phenomenological requirements.
For example,
it is possible to start from
``inaccessible'' textbook Hilbert space ${\cal L}$
and choose
the left-hand-side
``input information'' $\mathfrak{q}^{(JM)}$
as a diagonal
operator with equidistant spectrum simulating the position on
a
one-dimensional discrete lattice or on
its suitable continuous-spectrum limit \cite{Jones,Jonescomm}.

\section{Norm-ambiguity paradox and its consequences\label{violat}}

A concise outline of the
non-Hermitian but unitary
theory
of closed systems
is relocated to
Appendices A -- D.
Using the notation
of diagram (\ref{ufin})
in Appendix A let us now emphasize that in most
applications
the information about dynamics
is carried just by the
Hamiltonian $H$ acting in
an auxiliary
Hilbert space ${\cal K}$
in which $H \neq H^\dagger$.
As a consequence,
the choice of metric $\Theta$
compatible with the quasi-Hermiticity condition (\ref{zasdieu})
remains non-unique  \cite{Geyer}.
The
relevant (i.e., physical, $\Theta-$dependent)
size of perturbations $V$ in (\ref{perham})
is, therefore, indeterminate.

This is a paradox,
the relevance of which becomes particularly serious
in the realistic models of quantum systems in which
one cannot ignore the possible occurrence of
random, uncontrolled, statistically distributed perturbations.

\subsection{Random perturbations and pseudospectra}

In the most common textbook version of
quantum mechanics of the perturbed unitary systems living  in ${\cal L}$
the evolution
is generated by
the perturbed Hamiltonians
which are self-adjoint,
 $$
 \mathfrak{h}(\lambda) =\mathfrak{h}(0)+\lambda\,\mathfrak{h}_1=
 \mathfrak{h}^\dagger
 $$
(see \cite{Messiah}
or Eq.~(\ref{tarela}) in Appendix B).
The stability
of the system may be then tested
using all
perturbations,
the norm of which is bounded, $\|\mathfrak{h}_1\|\leq \epsilon$.
For this purpose the spectra of the
perturbed Hamiltonians
could be calculated using the $\lambda-$dependent
Schr\"{o}dinger equation in ${\cal L}$,
 \be
 \mathfrak{h}(\lambda)\,|\psi_n(\lambda)\pkt =
 E_n(\lambda)\,|\psi_n(\lambda)\pkt\,,
  \ \ \ \ \
 n = 0, 1, \ldots\,
 \label{selfadjo}
 \ee
plus, say,
the Rayleigh-Schr\"{o}dinger
perturbation-series ansatz (\ref{20}).

As a result one would obtain, in principle at least,
a union of all of the possible perturbed spectra,
i.e.,
the set
 \begin{equation}\label{RS-thm}
  \bigcup_{\lambda\,\|\mathfrak{h}_1\| <
  \epsilon} \sigma(\mathfrak{h}(0)+\lambda\,\mathfrak{h}_1)
  \,
 \end{equation}
which should lie, for stable systems, just inside a small
vicinity of $\sigma(\mathfrak{h}(0))$,
i.e., of the unperturbed spectrum.
In such a setting it is recommended to recall the
Roch's and Silberman's observation \cite{RST}
that the set (\ref{RS-thm}) coincides with the
so called pseudospectrum $\sigma_\epsilon(\mathfrak{h}(0))$
of $\mathfrak{h}(0)$, i.e., with the set
which is defined as the
following union of the spectrum
and of the domain in which the resolvent of $\mathfrak{h}(0)$
remains large \cite{Trefethen},
 \begin{equation}
 \label{pseudo}
 \sigma_\epsilon(\mathfrak{h}(0))= \sigma(\mathfrak{h}(0)) \cup
  \big\{
  z \in {\mathbb{C}} \ \big| \
  \|(\mathfrak{h}(0)-z)^{-1}\| > \epsilon^{-1}
  \big\}\,.
 \end{equation}
One can cite paper \cite{Viola} and conclude that
``if $\mathfrak{h}$ is self-adjoint \ldots'', the pseudospectra
``give no additional information''.


\subsection{Norms in non-Hermitian models}

Let us repeat that
as long as the Hamiltonians in question are kept self-adjoint,
the Roch's and Silberman's observation simplifies the
analysis of the influence of random perturbations
because it just shows that
the smallness of perturbations
immediately implies
that at the sufficiently small $\epsilon$
the difference between the sets $\sigma_\epsilon(\mathfrak{h}(0))$
(pseudospectrum)
and $\sigma(\mathfrak{h}(0))$ (spectrum) becomes negligible.

The situation becomes thoroughly different
when a quantum Hamiltonian
$H$ is chosen ``highly non-self-adjoint'' because then,
``the pseudospectrum
$\sigma_\epsilon(H)$
is
typically much larger than the $\epsilon-$neighborhood
of the spectrum''.
There is a subtlety in such a proposition
(cited from~\cite{Viola})
because in the
context of the general non-Hermitian Schr\"{o}dinger Eq.~(\ref{sekvoj})
one has to distinguish,
in a way already emphasized in Introduction, between
its open-system and closed-system interpretations.

In the former,
``resonances-describing'' subcase
we would have to complement
Eq.~(\ref{sekvoj})
by the specification of the conventional
Hilbert space
endowed with the usual, metric-independent
norm. In diagram (\ref{ufin})
such a space is denoted by the dedicated symbol ${\cal K}$,
with the norm of $V$ denoted as
$\|V\|$ as usual. Hence, in such a case
(not, by the way, of our present immediate interest),
we may formally set
$\Theta=I$ and ${\cal H}={\cal K}$ in Appendix A.

In the other,
``bound-states-describing'' subcase
(which {\em is\,} of our present interest)
we may still
follow the same
conventions
as introduced in Appendix A. Thus,
with $\Theta \neq I$ and with ${\cal H} \neq {\cal K}$
we have to treat
Schr\"{o}dinger  Eq.~(\ref{sekvoj})
as living in an amended,
physical Hilbert space
${\cal H}$.

Unless one asks questions
about norms, only the dual versions of
the vector spaces ${\cal K}$ and ${\cal H}$ are different.
Still,
precisely the difference between the
operator norm of $V$ in ${\cal K}$ (denoted as usual, $\|V\|$)
and in ${\cal H}$ (to be denoted differently, say, as $\sharp V\sharp$)
becomes one of the most essential aspects of the
respective alternative definitions of the
Hilbert-space-dependent pseudospectra.

\subsection{Pseudospectra in quasi-Hermitian models}

As long as we are not going to study resonances,
we may just restrict our attention to the random perturbations
in quasi-Hermitian (i.e., by definition, in the hiddenly unitary)
closed quantum systems.
In principle, their description in
the alternative physical Hilbert spaces ${\cal L}$ and  ${\cal H}$
is then equivalent.
In practice, nevertheless,
one may observe that the predictions
of the measurements as constructed
in the
textbook Hilbert-space representation
are impractical and less user-friendly.
Then we are forced to treat the knowledge of
the union  (\ref{RS-thm})
of the perturbed spectra in ${\cal L}$
as ``technically inaccessible''.

After we decide to move to ${\cal H}$
we must also remember that
the corresponding
physical norm $\sharp V\sharp$ of
perturbations
becomes different and, first of all,
$\Theta-$dependent.
The key and meaningful question to ask is then the
question
about the structure of
the union
 \begin{equation}\label{uRS-thm}
  \bigcup_{\lambda\sharp {V} \sharp\,
  < \epsilon} \sigma({}{H(0)}+\lambda\,{}{V})
  \,
 \end{equation}
of the spectra
of all of the slightly but
randomly perturbed  systems living
in  ${\cal H}$.

The above-cited theorem
can be recalled again.
After one defines the
pseudospectrum
 \begin{equation}\label{upseudo}
  \sigma_\epsilon({}{H}) := \sigma({}{H}) \cup
  \big\{
  z \in {\mathbb{C}} \ \big| \
  \sharp ({}{H}-z)^{-1} \sharp\, > \epsilon^{-1}
  \big\}\,
  \,
\end{equation}
in ${\cal H}$, one immediately obtains
the Roch-Silberman relationship
 \begin{equation}\label{vuRS-thm}
  \bigcup_{\lambda\sharp {V} \sharp\,
  < \epsilon} \sigma({}{H(0)}+\lambda\,{}{V})
  =\sigma_\epsilon({}{H(0)})
  \,.
 \end{equation}
This is our desired ultimate formula.
In the correct and physical
Hilbert space  ${\cal H}$
in which the Hamiltonian is made self-adjoint,
this formula defines the sensitivity to perturbations
in terms of
the correct physical pseudospectrum (\ref{upseudo}).
Its explicit numerical construction
is facilitated
and made useful.
Obviously, once we require our random perturbations to be small
in ${\cal H}$,
we may again
recall Proposition Nr. 3 in \cite{Viola}
and
conclude that
in full parallel with the Hermitian models also
in the quasi-Hermitian picture of dynamics
the spectrum and
pseudospectrum
carry equivalent information about the
sensitivity of bound states to perturbations,
 \be
  \sigma_\epsilon(H(0))
  \subseteq
  \big\{z\in {\mathbb{C}}\ \big| \ {\rm dist\,}\big(z,\sigma(H(0))\big)
   < const\,\times \epsilon \big\}
  \,.
  \label{trivi}
 \ee
At the small
values of $\epsilon$
the pseudospectrum
is formed just by a small vicinity
of the spectrum.
In the terminology of paper \cite{Viola}
such a pseudospectrum is ``trivial''
because
small random perturbations
cannot destroy the stability of
the underlying closed quantum system.

\section{Amended Rayleigh-Schr\"{o}dinger construction\label{whava}}

Let us temporarily return to
the open-system theory where one does not need
to define any nontrivial inner-product metric
because
the evolution is non-unitary (cf., e.g.,
monographs \cite{Horacek} or \cite{Nimrod}).
In section \ref{aaaa}
we recalled, for illustration,  paper \cite{BD}
as a typical sample
of such a more traditional approach.
Bender with Dunne used there
a Hamiltonian~(\ref{1})
for the purposes of
description of a
complicated physical phenomenon.
The physical
Hamiltonian itself,
as sampled by Eq.~(\ref{physi}), has only been
deduced after an
analytic-continuation redefinition
of the model.

In our present paper our strategy is different,
with our attention restricted to the unitary,
closed and
stable quantum systems in which the
unitarity of evolution
of evolution coexists with the
non-Hermiticity of $H$.
In this setting we intend to describe an amendment of the QHQM
perturbation-expansion recipe
in which the metric-related technical obstacles
will be circumvented using a
reformulation of the theory as recently
proposed, in different context, in \cite{twoSE}.


\subsection{The choice-of-space problem revisited\label{bbb}}

The requirement of unitarity of the evolution
may make the QHQM perturbation theory
discouragingly complicated,
mainly due to the operator-expansion nature of
the newly emerging series (\ref{2x}) representing the metric.
In a way outlined in Appendix C,
the theory has to be formulated in as many as five Hilbert spaces
(cf. our present diagram (\ref{5ufin})
or
an analogous diagram Nr.~(20) in \cite{response}).
The
standard, reference-providing space ${\cal L}$ of textbooks
has to be accompanied by
the doublet of the preferred representation spaces,
viz., by ${\cal K}(\lambda)$ pertaining to the ultimate dynamical
scenario  and by ${\cal K}(0)$ representing the
solvable unperturbed system.
The remaining pair of their
amended physical partners
consists of the
predictions-offering
${\cal H}(\lambda)$ (carrying the ultimate
picture of physics) and ${\cal H}(0)$ (i.e., its unperturbed
$\lambda = 0$ partner).

The five-Hilbert-space pattern looks complicated.
Concerning its applicability, one has to be a bit skeptical.
In what follows we intend to show that
a fairly efficient remedy of the skepticism
can be based on a
more or less straightforward reformulation of the theory
in which the specification of the metric
will be re-interpreted as an upgraded form of
transition from ${\cal K}$
to the correct physical Hilbert space ${\cal H}$.
A motivation of our present modification of the theory
lies in an undeniable appeal of the
Rayleigh-Schr\"{o}dinger perturbation-approximation
philosophy which may be characterized,
in the conventional textbook setting, by
its enormous technical simplicity.
In this sense, we intend to show that
this simplicity need not get lost
after one moves to
the innovative QHQM framework.

Our attention will be concentrated
upon the mathematical consistency aspects of the theory.
We will emphasize that it is possible to overcome
the most
unpleasant conceptual complications
emerging when
one deals with
a realistic quasi-Hermitian Hamiltonian
of a unitary quantum system which is allowed
to vary with a parameter.
The theory will be re-analyzed in a way
inspired by several publications
a sample of which is recalled
in Appendix~C.

Attention will be paid to
the models in which the parameter-dependence
remains weak and tractable by the
techniques of perturbation theory \cite{Kato}.
In the first step of amendment
of the conventional approaches
we will modify the very concept of a state,
keeping in mind that
in conventional textbooks, the state is usually characterized
by a ket-vector element 
of a physical Hilbert space (i.e., by $|\psi\kt \in {\cal H}$).
The most immediate inspiration of a change of such a definition
may be deduced from Eq.~(\ref{melasa})
in which it is sufficient to abbreviate
 \be
 \br \psi(\lambda)|\Theta(\lambda)
 :=\bbr \psi(\lambda)|\ \in \ {\cal K}'
 \,
 \label{ketizonb}
 \ee
or, after the Hermitian conjugation in
our mathematical representation space,
 \be
 \Theta(\lambda)| \psi(\lambda)\kt
 :=| \psi(\lambda)\kkt\ \in \ {\cal K}
 \,.
 \label{ketizonc}
 \ee
These abbreviations enable us to rewrite Eq.~(\ref{melasa})
in a more compact form
 \be
 a(\lambda) =
 \bbr \psi_n(\lambda)| A |\psi_n(\lambda)\kt
 \label{melasab}
 \ee
out of which the metric $\Theta(\lambda)$ seems to have ``disappeared''.

An easy resolution of such an apparent paradox is that
we moved back from auxiliary ${\cal K}$
to physical ${\cal H}$.
After some elementary algebra we also reveal that
the parallels between the ``old''
ket vectors $ |\psi_n(\lambda)\kt \in {\cal K}$ and
their ``new'' partners of Eq.~(\ref{ketizonc})
(which could be called ``ketkets'')
can even be extended yielding an identically satisfied ``parallel''
eigenvalue problem
 \be
 H^\dagger(\lambda)\,|\psi_n(\lambda)\kkt =
 E_n(\lambda)\,|\psi_n(\lambda)\kkt\,,
  \ \ \ \ \
 |\psi_n(\lambda)\kkt \in {\cal K}_{} \,,
 \ \ \ \ \
 n = 0, 1, \ldots\,
 \label{2nonselfadjo}
 \ee
(with the same real spectrum of course) or,
after the mere Hermitian conjugation in ${\cal K}$, equivalently,
 \be
 \bbr \psi_n(\lambda)|\,H(\lambda) =
 \bbr \psi_n(\lambda)|\, E_n(\lambda)\,, \ \ \ \ \
 \bbr \psi_n(\lambda)| \in {\cal K}'_{} \,,
   \ \ \ \ \
 n = 0, 1, \ldots\,.
 \label{Lnonselfadjo}
 \ee
Now, we are prepared to realize that for vectors,
the  ``physical'' Hermitian conjugation as defined, hypothetically,
in the ``hidden'' Hilbert space ${\cal H}$ just replaces
the ket
$|\psi_n(\lambda)\kt \in {\cal H}$
by the ``brabra'' $\bbr \psi_n(\lambda)| \in {\cal H}'$.

Summarizing,
we come to the conclusion that
in the correct physical Hilbert space ${\cal H}$
the most natural representation of an $n-$th bound state
of the quantum system in question
will not be provided by any ket but rather by the elementary projector
 \be
 \varrho_n(\lambda)=|\psi_n(\lambda)\kt
 \frac{1}{\bbr \psi_n(\lambda)|\psi_n(\lambda)\kt}
 \bbr \psi_n(\lambda)|\,.
 \label{densia}
 \ee
The main advantage of such an upgrade of conventions is twofold.
First, formula (\ref{densia}) remains the same in both
of the Hilbert-space representations in
${\cal K}$ and in ${\cal H}$, and second,
using the standard
definition
 \be
 a_n(\lambda)={\rm Tr}[A \varrho_n(\lambda)]\,
 \label{trejs}
 \ee
of the probability density,
one immediately rediscovers the above-mentioned
equivalent measurement-predicting prescription (\ref{melasa}).
Moreover, the use of formula also opens the way from pure states
to mixed states and quantum statistical physics \cite{NIPa,NIP,NIPb}.

\subsection{Rayleigh-Schr\"{o}dinger construction revisited}


In the light of our preceding considerations
the essence of our present innovation
of the QHQM Rayleigh-Schr\"{o}dinger construction
of the series (\ref{20}), (\ref{RSer})
and (\ref{2x})
[with an implicit reference to
the ``measurement-prediction'' formula (\ref{melasa})
{\it alias\,} (\ref{melasab})
{\it alias\,} (\ref{trejs})]
can be seen to lie simply in the
replacement of the almost prohibitively complicated
operator-expansion formula (\ref{2x})
by the alternative and formally sufficient
new ketket-expansion ansatz
 \be
 |\psi(\lambda)\kkt =|\psi(0)\kkt + \lambda\,|\psi^{(1)}\kkt
 +\lambda^2\,|\psi^{(2)}\kkt+\ldots\,.
 \label{RSlef}
 \ee
In other words, we will still have to
start from the entirely conventional
decomposition (\ref{perham}) of the Hamiltonian
and from the related order-by-order re-arrangement
 \be
 \left [H - E(0)
  + \lambda\,({V}-E^{(1)}) -
 \lambda^2\,E^{(2)}-\ldots
  \right ]
  \left [\,|0\kt + \lambda\,|\psi^{(1)}\kt
 +\lambda^2\,|\psi^{(2)}\kt+\ldots \right ]=0\,
 \label{27ap}
 \ee
of our initial perturbed form of
Schr\"{o}dinger equation (\ref{sekvoj}).
The innovation only comes when we
reject the recipe of our previous proposal \cite{response}
[based on the reconstruction of $\Theta(\lambda)$
via the clumsy power-series ansatz  (\ref{2x})]
as unnecessarily (and, what is worse, more or less
prohibitively) complicated.

In our present upgraded recipe one
simply complements Eq.~(\ref{27ap})
by its associated partner for ketkets,
 \be
 \left [H^\dagger - E(0)
  + \lambda\,({V^\dagger}-E^{(1)}) -
 \lambda^2\,E^{(2)}-\ldots
  \right ]
  \left [\,|0\kkt + \lambda\,|\psi^{(1)}\kkt
 +\lambda^2\,|\psi^{(2)}\kkt+\ldots \right ]=0\,.
 \label{27apk}
 \ee
Obviously, an enormous simplification of the
construction of the
measurable
predictions (\ref{trejs}) is achieved.
Indeed, in comparison with the complicated formulae
of paper \cite{response}
teh construction of the necessary
recurrences for the sequence of corrections
becomes immediate, making just use of the
slightly upgraded
projector
 \be
 \Pi= I - |0\kt \bbr 0|=\sum_{j>0}\,|j\kt \bbr j|\,
 \ee
and leading to the easily deduced formulae for the energies, say,
 \be
 E^{(1)}=\bbr 0|{V}|0\kt\,,\ \ \
 E^{(2)}=\bbr 0|{V} \Pi |\psi^{(1)}\kt\,,\ \ \ \ldots\,
 \label{40}
 \ee
as well as to the kets
 \be
 |\psi^{(1)}\kt = \Pi\,\frac{1}{E(0)-\Pi H\Pi}\Pi {V} |0\kt\,,
 \ee
 \be
 |\psi^{(2)}\kt = \Pi\,\frac{1}{E(0)-\Pi H\Pi}\Pi [{V}-E^{(1)}]\Pi |\psi^{(1)}\kt\,,
 \ee
(etc) and, analogously, for the ketkets,
 \be
 |\psi^{(1)}\kkt = \Pi^\dagger\,\frac{1}{E(0)
 -\Pi^\dagger H^\dagger\Pi^\dagger}\Pi^\dagger {V^\dagger} |0\kkt\,,
 \ee
 \be
 |\psi^{(2)}\kkt = \Pi^\dagger\,\frac{1}{E(0)
 -\Pi^\dagger H^\dagger\Pi^\dagger}\Pi^\dagger
 [{V^\dagger}-E^{(1)}]\Pi^\dagger |\psi^{(1)}\kkt\,,
 \ee
etc.

Summarizing, one only has to remind the readers that
the full-fledged version of the present
amended QHQM perturbation theory is only needed when
we really have to predict the results of measurements
of the observable represented by a preselected operator $A$.
In applications, we are often interested just in the evaluation of
just
one of the values of the energy (which is, moreover,
defined as one of
the eigenvalues of the Hamiltonian).
In practice, such a value is often known to be real.
In such a case, naturally,
what is needed is just the more or less standard
construction of the single
power series (\ref{20}).
We may conclude that precisely such simplified
calculations were performed in
papers \cite{Rafa} and \cite{BD},
with the details recalled in section \ref{aaaa}
and in subsection \ref{baa}
above.

\section{Discussion\label{onintrod}}



\subsection{Key role played by the proof of
reality of spectrum\label{fxbb}}

In the early studies
of non-Hermitian Hamiltonians
with real spectra \cite{BB,BG,BM,Caliceti}
the authors
admitted
that the
non-Hermiticity
of $H(\lambda)$
could make the standard probabilistic closed-system interpretation
of the states questionable.
For example,
Bender with Dunne \cite{BD}
circumvented the problem by
claiming that their
expansion (\ref{2})
just offers a
``strong evidence'' that the quantity $E(\lambda)$
is an analytic function
which can be continued to the cut complex plane of
couplings $g=\lambda^2$.

Later,
emphasis has been shifted to the
requirement of the reality
{\it alias\,} potential observability of the would-be
bound-state energy-level spectrum of $H$
representing
a necessary condition of
existence of an amended inner product.
A direct and truly innovative
closed-system physical
interpretation of models
started to be sought in the reconstruction of metric
$\Theta=\Theta(H)$ \cite{Carl,ali}.

In the context of QHQM perturbation theory,
for several reasons (some of which have been discussed above),
the necessity of the proof of the reality of the energy spectra
also acquired a new urgency.
In its analysis
as performed in our preceding paper \cite{response}
we emphasized that the scope of the QHQM perturbation theory
is in fact ``too broad''.
In comparison with the
constructive strategy of conventional textbooks
(where the trivial physical inner-product metric is
chosen in advance),
the more flexible QHQM theoretical framework
forced us to admit that our $\Theta$
must be treated as perturbation-dependent.
The two conventional  Rayleigh-Schr\"{o}dinger power series (\ref{20})
and (\ref{RSer}) had to be complemented
by the third item (\ref{2x})
representing the metric
and making the construction of the model
(i.e., of its correct physical Hilbert space)
almost prohibitively difficult.

In this context, one of our present main results is that
we managed to simplify the construction
by replacing the difficult operator expansion (\ref{2x})
by its mere ketket-vector alternative (\ref{RSlef}).
Nevertheless, even after such an upgrade of the recipe
the (rarely easy!)
proof of the reality of the spectrum will
still keep playing the most important role of
a necessary preparatory step in applications.


\subsection{The requirement of completeness of the set of observables}

We
achieved a
simplification of the non-Hermitian version of the
Rayleigh-Schr\"{o}dinger
formalism by
making the operator-expansion (\ref{2x}) of the metric ``invisible''.
The price to pay was
the loss of insight in the
correspondence between the
reality of spectrum
and the choice of the class of admissible perturbations.
In fact, the study of this correspondence is
nontrivial requiring, probably, a return to the study of
explicit expansions (\ref{2x}).

The question
remains to be
kept in mind as a truly interesting and challenging
future project, nevertheless.
One of the reasons is that
it is closely related
to the
paradox of the ambiguity of the metric.
Indeed, it is well known
that the operator
$\Theta$
endowing a given Hamiltonian $H$
with a self-adjoint status in ${\cal H}$
need not be unique.
As a consequence,
even the norm of perturbation
$V$
in ansatz (\ref{perham})
can vary so that also
the conventional condition of its ``sufficient smallness''
could be difficult if not impossible to prove.

The ambiguity of $\Theta=\Theta(H)$
has been identified, in review \cite{Geyer},
as resulting from an incompleteness of our information about
the system's dynamics. Indeed, the emergence of any independent
candidate $\Lambda$ for an observable
(which would have to be quasi-Hermitian with respect to the same metric,
$\Lambda^\dagger\Theta=\Theta\,\Lambda $)
would suppress the ambiguity of $\Theta$ whenever
such a candidate appears not to be reducible to a function of  $H$,
$\Lambda \neq \Lambda(H)$.
This means that a unique $\Theta$ will be obtained
only after one specifies a complete set of irreducible
observables $H \ (=\Lambda_0)$ and $\Lambda \ (= \Lambda_1)$
and, perhaps, $\Lambda_2$ etc
\cite{Geyer}.

In a way discussed in paper \cite{Geyer}
one is usually forced to work just with
an incomplete irreducible set of preselected observables $\Lambda_j$.
This means that the ambiguity of the metric
can only rarely be fully suppressed.
One may try to circumvent the problem
by making a more or less arbitrary choice of
one of the eligible metrics.
The same strategy is, after all, widely accepted
in the conventional textbooks
using trivial $\Theta=I$.

In the framework of unconventional QHQM,
an exhaustive explanation of the
problem
of the ambiguity of the norm
can already be found discussed in review \cite{Geyer}
where one reads that the variability of our choice of the metric
just reflects an incompleteness of the input information
about dynamics.  This means that such an ambiguity
disappears when our knowledge of $H$ becomes
complemented by the knowledge of a
sufficiently large
(i.e., in mathematical language, ``irreducible'')
set of some further operator
candidates
for the observables.

In this sense we arrive at a new paradox.
Either we postulate such a knowledge or not.
Naturally, the abstract theory
would be only fully satisfactory in the former case.
In such a case, nevertheless,
the $\lambda-$dependence of the
Hamiltonian and metric
would be inherited by an induced
and strongly counterintuitive $\lambda-$dependence of
all of the further (i.e., necessarily quasi-Hermitian)
observables
$\Lambda_j$ with $j>0$.

\subsection{The coordinate-non-observability paradox}

Among all of the
differential-operator candidates
for a closed-system quantum
Hamiltonian possessing a real energy-like spectrum
as sampled by Eq.~(\ref{1}) above,
one of the most interesting
alternative models was studied by
Buslaev with Grecchi \cite{BG}.
One of the truly striking features of their model
(which made it qualitatively different from Eq.~(\ref{1}))
was that for the
purposes of its mathematical consistency it was necessary
to keep the ``coordinate'' complex (i.e.,
$x \notin {\mathbb{R}}$,
in the asymptotic domain at least).
This is a contradictory situation because
such a variable cannot in fact be interpreted as
an observable quantity.

The puzzle has been clarified by an explicit
reference to
perturbation theory
in combination with the techniques
of analytic continuation.
In a way discussed also in section~\ref{aaaa}
above,
Buslaev with Grecchi
revealed
a hidden,
perturbation-series-mediated connection
between their manifestly non-Hermitian
``complex-coordinate'' oscillator
and the safely
self-adjoint
Hamiltonian
 \be
 {\mathfrak{h}}^{(AHO)}=-\triangle+|\vec{r}|^2+\lambda\,|\vec{r}|^4\,
 \label{opera4}
 \ee
describing an entirely conventional
quartic anharmonic oscillator \cite{Turbiner,Simon}.
They
were aware of the divergence of
the related
Rayleigh-Schr\"{o}dinger perturbation series (\ref{20}), but
their analysis
revealed the existence of an intimate relationship
between operator (\ref{opera4})
[defined as self-adjoint in the most common physical
Hilbert space  $L^2(\mathbb{R}^d)$]
and its specific non-Hermitian isospectral
descendant.

In paper \cite{BG} the
same idea has been shown to work also in application to
another, multiparametric
multiplet of
ordinary differential
Hamiltonian-like operators $H_n^{(BG)}$
with
$n=1,2,\ldots,K$ (with, incidentally, $K=7$).
A special status has been again enjoyed by the
element
 $
 H_1^{(BG)}={\mathfrak{h}}^{(BG)}
 $ which was required, in
the most conventional Hilbert
space $L^2(\mathbb{R})$,
self-adjoint.
The last element ${H}_K^{(BG)}$ of the sequence
appeared to be non-Hermitian but
parity-time-symmetric {\it alias\,}
${\cal PT}-$symmetric.
For our present purposes we may abbreviate
${H}_K^{(BG)} ={H}^{(BG)}$ (i.e.,
drop the last subscript)
and notice that
the above-mentioned
Jones' and Mateo's isospectrality relationship
finds a direct analogue in formula
 $$
 {\mathfrak{h}}^{(BG)} \sim H^{(BG)}.
 $$
This is not too surprising
because
the Jones' and Mateo's
Hamiltonian ${\mathfrak{h}}^{(JM)} $
is just a parameter-free special case
of the Buslaev's and Grechi's multiparametric
operator ${\mathfrak{h}}^{(BG)}$.
Thus, after
a multiparametric generalization
of
the Jones' and Mateo's
Dyson operator (\ref{[19dm]})
a
new light could be thrown
upon the concept of locality in
non-Hermitian physics
(cf. \cite{Liu} and also formula
(\ref{tarelax}) in section \ref{moco} above).

Incidentally,
the Jones' and Mateo's ``direction of
simplification'' becomes inverted since
the evolution controlled by $ H^{(BG)}$
has to be reclassified as a more complicated
picture
of dynamics.
Still,
the message which survives is that
the physical interpretation is directly
provided by the
Rayleigh-Schr\"{o}dinger perturbation series of Eq.~(\ref{20}).

\subsection{A detour to meaningful complex spectra}

In the conventional applications of
perturbation theory
one starts from the knowledge of
a preselected family of Hamiltonians
 \be
 H(\lambda)=H_0+\lambda\,H_1\,
 \label{elfadjo}
 \ee
in which the unperturbed operator $H_0$
is assumed maximally user-friendly or even,
often, diagonal.
The
specification of
the admissible
perturbations $\lambda H_1$ is then
rather routine, made in accordance with both the
phenomenological and mathematical model-building
needs \cite{Kato}.
Family (\ref{elfadjo})
is chosen,
in most textbooks,
as a mere sum of two self-adjoint
operators.

We have already emphasized
that
once one admits
a manifest non-Hermiticity of one or both of the operator
components of the Hamiltonian in an auxiliary Hilbert space
${\cal K}$,
 \be
 H_0 \neq H_0^\dagger\,,\ \ \ \
 H_1 \neq H_1^\dagger\,
 \label{lossoh}
 \ee
the technical costs of such a
weakening of the conventional assumptions
may be high
(cf. \cite{Carl,Geyer,ali,book,SIGMA,SIGMAb}).
Even when one decides
to keep
the working Hilbert space perturbation-independent,
${\cal K}(\lambda)={\cal K}(0)= {\cal K}$,
a number of challenging questions survives.
One of the most important ones follows from the possible
loss of the reality of eigenvalues,
 \be
 E(\lambda)=E(0)+\lambda\,E^{(1)}+ \lambda^2\,E^{(2)} + \ldots
 \in {\mathbb{C}}\,.
 \label{coan}
 \ee
Then,
one has to accept the open-system philosophy and
to treat
the Rayleigh-Schr\"{o}dinger
expansions
just as an ansatz which could work
even when $E(\lambda) \notin {\mathbb{R}}$
and even when the series is divergent.

The feasibility of such an alternative model-building strategy
has been confirmed, e.g.,
by Caliceti et al \cite{Caliceti}
(cf. also a more recent review of the field in \cite{CGbook}).
In essence,
the latter authors revealed
that in a number of specific toy models
the conventional ansatz (\ref{coan}) may still serve as a productive
constructive tool. Yielding,
at the small and real coupling constants,
the real (i.e., energy-like)
as well as complex (i.e., resonance-representing)
low-lying spectra
after standard resummation.

\subsection{Real spectra and the paradox of emergent instabilities\label{emerg}}

The reality of spectra of the Hamiltonians
has independently been noticed
in the context of quantum field theory
\cite{BM}.
This
attracted
attention of physics community
to the applicability of 
expansions (\ref{coan}) in the non-Hermitian setting of Eq.~(\ref{lossoh}).
The authors of the innovated studies of
imaginary cubic anharmonic-oscillator
Hamiltonians
 \be
 H^{(CAHO)}(\mu,\nu) = -\frac{d^2}{dx^2}+\frac{\mu^2}{4} x^2
 +{\rm i}\nu x^3\,
 \label{1re}
 \ee
identified
either $\lambda=\mu$
(say, in the ``strong-coupling expansions'' of Ref. \cite{Rafa})
or $\lambda=\nu$ (say, in the ``weak-coupling expansions''
of Ref. \cite{BD}).

Even when having the strictly real bound-state-like spectra,
the latter model-building efforts were
criticized by mathematicians
\cite{Trefethen,Viola}. They recommended a
replacement of the mere search for
eigenvalues (characterized as ``fragile'')
by a more ambitious construction of
pseudospectra.
We have to point out
that the
mathematically well founded latter
recommendation has been based
on a conceptual misunderstanding.
Fortunately, a disentanglement of the misunderstanding
was straightforward.
It proved sufficient to distinguish between the closed and
open systems
and to show that the construction of pseudospectra
only makes sense
(and offers new information)
in the latter case (for more details see also
section \ref{violat} above).

From a purely mathematical point of view
one should not be too much surprised by the
latter conclusion and, in particular,
by the ``wild'' \cite{Viola}
behavior of open systems
exhibiting emergent instabilities
because the
theory behind the closed systems is different.
For them, the
constructions and predictions obtained
in
the two
alternative ``physical'' Hilbert spaces ${\cal L}$ and ${\cal H}$
are, by definition, equivalent.
Thus, no paradox can be seen in the
existence of the mechanism due to which
the pseudospectra of closed systems remain
well behaved even when the
representation of dynamics itself is non-Hermitian.

Another, even more straightforward explanation
of the existence of the
emergent open-system instabilities
becomes best visible
when the system in question happens to
lie close to a Kato's exceptional-point singularity \cite{Nimroda,Nimrode,EP3}.
In such a vicinity, indeed, the operator of metric $\Theta$
becomes singular and dominated by a projector \cite{EPsa}.
Perturbations $H_1$ which are small
with respect to the correct physical
Hilbert-space norm (in our present notation
this means that
$\sharp H_1\sharp \ll 1$)
may still be, simultaneously, very large with respect
to the conventional open-system norm as defined in the
standard regular limit of
$\Theta \to I$ (i.e., $\|H_1\| \gg 1$).
As a consequence,
perturbations
may be expected to lead
to the ``wild'' forms of pseudospectrs (\ref{RS-thm})
as sampled, via a number of elementary
examples, in \cite{Viola}.

\subsection{Ultimate challenge: Models where the metric does not exist\label{bxbb}}

To a compact introduction in the overall QHQM theory
as provided in Appendix A
it makes sense to add that
a truly enormous increase of popularity
of the formalism has been inspired
by the Bender's and Boettcher's
claim \cite{BB} that the reality of spectra
is a phenomenon which can be observed in an unexpectedly
broad class of models which are
not only phenomenologically attractive
but also mathematically user-friendly.
These
results
set the scene for an intensive subsequent study.
No surprise: Whenever
an operator $H$ proves non-Hermitian (in ${\cal K}$) while its
spectrum $\{E_n\}$ is ``bound-state-like''
(i.e.,
real, discrete and bounded from below),
one feels tempted to consider
the possibility of its quantum quasi-Hermitian Hamiltonian-operator
interpretation.

In \cite{BB} the temptation has been further supported
by the detailed
analysis of the specific
ordinary differential Hamiltonian-like operators
 \be
 H^{(BB)}(n) =  -\frac{d^2}{dx^2}   
 - ({\rm i} x)^{n+2}\,,\ \ \ \ \ n\geq 0\,.
 \label{ante1}
 \ee
These operators are, in general,
complex and manifestly non-Hermitian
but still possessing the strictly real
and discrete bound-state-like
spectra. On these grounds
Bender with Boettcher conjectured that
such operators could be treated as
Hamiltonians
in certain unconventional, ``analytically continued'' quantum theories.

In 2012,
Siegl with Krej\v{c}i\v{r}\'{\i}k \cite{Siegl}
opposed the claim.
Using the rigorous methods of functional analysis they proved
that for
at least some of the
toy models $H$ having the elementary
differential-operator form (\ref{ante1})
an acceptable metric $\Theta$
which would satisfy relation (\ref{zasdieu})
does not exist
at all.
This weakened the enthusiasm because
at least some of the local-interaction benchmark models
cannot be endowed with {\em any\,}
admissible
physical-Hilbert space  ${\cal H}$.

One of the ways of circumventing such a mathematical
disproof of quasi-Hermiticity has been
found in
a transition to the
open system reinterpretation of the models \cite{Christodoulides}.
As a  benefit, such a change of strategy
simplified the mathematics because
one could simply set $\Theta=I$.
A return to the old open-system philosophy
behind models (\ref{ante1})
appeared even productive in mathematics:
In a way outlined in section \ref{emerg} above
it led to
the discovery of certain unexpected
spurious
approximate solutions of Eq.~(\ref{sekvoj})
emerging
at the
energies which lied far from the spectrum \cite{Viola}.
Thus,
it was immediate to conclude that in place of the spectrum,
the much more useful descriptive tool
can be sought in the pseudospectra.

In a way which we described in section~\ref{violat} above, the
pseudospectra directly
characterize the influence of random perturbations
upon dynamics of the systems. Incidentally,
their analysis has been shown to make sense only in
the open-system cases in which the
spectra of $H$ are not real.
As long as ${\rm Im\, }E_n \neq 0$ at some $n$,
the Hamiltonian cannot be Hermitian, $H \neq H^\dagger$.
Thus, we may set $\Theta=I$ and identify ${\cal H}={\cal K}$.
In contrast, the knowledge of pseudospecta is not neeeded in
the other, stable-bound-state scenario, partially because
the implication
 $
 [{\rm Im\, }E_n \neq 0]\ \Longrightarrow \ [H \neq H^\dagger]
 \,
 $
cannot be inverted.

It is possible to conclude that
for the closed systems
with the
Hamiltonians sampled by Eq.~(\ref{ante1}), the question of their
acceptability is still open. The reason is that
the physical Hilbert space
defined in terms of the correct inner-product metric $\Theta=\Theta(H)$
need not exist.
Thus, whenever we decide to stay
inside the QHQM theory and
require that
 $$
 [H \neq H^\dagger]\ \& \ [{\rm Im\, }E_n = 0]\,,\ \ \forall n
 \,
 $$ 
we must keep in mind that
the status of
many popular
illustrative
examples
has to be reconsidered as inconclusive,
with an acceptable physical
interpretation
being still sought in several new
directions \cite{Makris,Connes,SKbook,Uwe,denish,gride,ozky,Joshua,Semoradova}.

In fact, already Scholtz et al \cite{Geyer},
were probably aware of the similar
mathematical subtlety because
they complemented
the quasi-Hermiticity requirement (\ref{zasdieu})
by a few
further
consistency-supporting sufficient conditions.
Among them, the
most prominent amendment of the theory seems to be their mysterious
requirement of the boundedness of $H$. Unmotivated and
counterintuitive as it might have looked in the past,
it was probably one of the lucky
parts of the formulation of QHQM by Scholtz et al \cite{Geyer}
because,
in retrospective, it
excludes the contradictory differential unbounded-operator
models (\ref{ante1})
as unacceptable.

\section{Summary\label{xonintrod}}

In the textbooks
on quantum theory the authors have to distinguish
between the models supporting, and not supporting,
the presence of resonances.
The notion of perturbation
plays a fundamental role in both of these implementations of
the theory.
For two reasons. The first one is realistic:
Whenever one
tries to prepare and study a quantum system, stable or unstable,
it is hardly possible to achieve
its absolute isolation from an uncontrolled
environment.
One has to guarantee the negligibility of
influence of such an environment
using, typically,
non-Hermitian Hamiltonians and
open-system models with complex spectra and random perturbations.

The
second reason is mathematical:
Even if we manage to guarantee that
the system in question is, up to negligible errors,
isolated,
perturbation theory re-appears as a powerful tool
suitable for calculations and
for an efficient evaluation of predictions.
Naturally, a
consistency of perturbation-related constructions
needs a guarantee of a ``sufficient
smallness'' of the perturbation.
Such a guarantee is a
task, an explicit formulation of which
depends on the
model-building details. Our present attention
has mainly been devoted, therefore, to the
physics
of stable bound states
(and just marginally to unstable resonances)
in a way motivated by
the recently increasing popularity
of
the so called
non-Hermitian Schr\"{o}dinger representations
of the stable and unitary quantum systems.

In the literature the presentation of this subject
may be found accompanied
by the emergence of multiple new and unanswered questions.
In our paper we picked up a few of such questions
which we were able to answer. Basically, our answers
may be separated in several groups.
In the first one we felt inspired by the
authors who studied the pseudospectra.
We imagined that in such an area of
research the application of innovative mathematics
is not always accompanied by a clear explanation of physics.
In this setting we conjectured that
a key to the resolution of certain emerging apparent paradoxes
can be found in distinguishing, more consequently,
between the
traditional non-Hermitian quantum models with $\Theta=I$
(for the study of which the pseudospectra have been found truly
indispensable \cite{Trefethen})
and the more recent and sophisticated closed-system theories in which the
specification of the correct inner-product metric proves nontrivial and
Hamiltonian-dependent,
$\Theta =\Theta(H) \neq I$.

In this way we managed to explain that
in certain applications (sampled by the random-perturbation studies)
there is practically no difference between the
use of the QHQM and standard quantum mechanics.
A slightly different conclusion has been achieved
when we turned attention to a more explicit study of the
Rayleigh-Schr\"{o}dinger perturbation expansions.
The differences, not too well visible in the mere calculations of
energies \cite{Rafa,BD}, appeared immediately deeply relevant
when one becomes interested in practically any other observable
quality/quantity of the system.

The latter feature of the theory
has already been observed (and not found too welcome) in
our preceding paper \cite{response}.
Our
detailed analysis of the structure of the QHQM-version of perturbation
theory led us there to a few rather skeptical conclusions concerning the
applicability of the formalism in its full generality.
In our present paper we managed to show that the
strength of the latter discouraging results
can perceivably be weakened when one
reconsiders the theory and after one reduces
its scope just to
the description of
its experimentally verifiable predictions.

In this spirit we proposed to replace the
next-to-prohibitively difficult operator-valued solution
of Eq.~(\ref{zasdieu})
(specifying the perturbation-dependent
metric $\Theta(\lambda)$
needed in Eq.~(\ref{melasa}))
by the vector-valued solution
of Eq.~(\ref{2nonselfadjo})
entering the modified form (\ref{melasab}))
of the same prediction which is, even by itself, much easier
to evaluate.

In conclusion it is probably worth adding that
along the same methodical lines one could also get beyond the framework
of the Schr\"{o}dinger picture in which
the opereators of observables are mostly assumed
time-independent,  $A\neq A(t)$.
In the future, perhaps, the same methodical ideas
might prove applicable also in the
non-stationary context and models and in the
interaction-picture
extension of the hiddenly unitary-evolution formalism
as proposed, couple of years ago, in \cite{timedep}.

\newpage

\section*{Appendix A: Quantum mechanics
in quasi-Hermitian representation\label{ec2}}

A comprehensive outline of the formulation of unitary quantum mechanics
in which the conventional requirement of
Hermiticity of the Hamiltonian is replaced by an
apparently weaker, metric-dependent
quasi-Hermiticity constraint (\ref{zasdieu})
can be found not only in the older review by Scholtz
et al.~\cite{Geyer} but also in
a few
newer
papers (e.g., \cite{Carl,ali,SIGMAdvab}) and
books (e.g., \cite{Carlbook,book}).
In the interpretation of review \cite{SIGMA}
the formalism
is based on
simultaneous use of a triplet of Hilbert spaces
(say, $[{\cal L},{\cal K},{\cal H}]$)
connected by the Dyson-inspired \cite{Dyson}
mutual correspondences as displayed in the following diagram,
%
 \be
  \ \ \ \ \ba
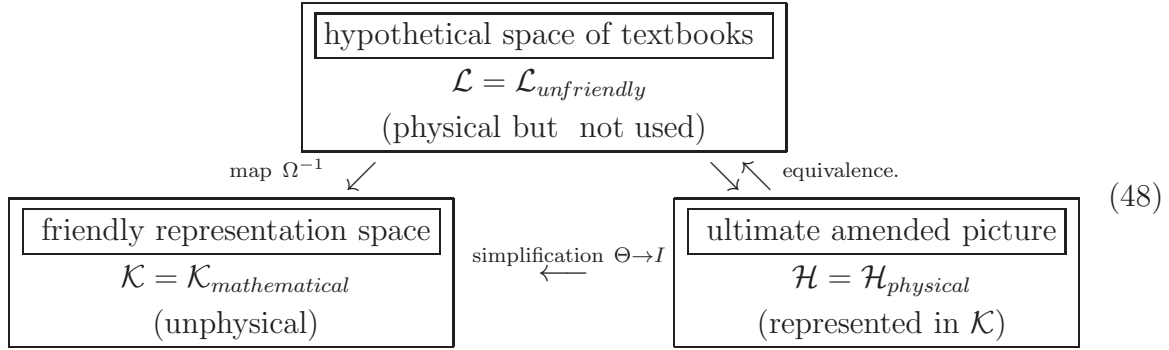
    \begin{array}{|c|}
 \hline
 \vspace{-0.5cm}\\
  \fbox{\rm  {hypothetical\ space\ of\ textbooks
    }}
  \\
  \ {\cal L}= {\cal L}_{un\!friendly}
 \\ {\rm (physical\ but\ \ not\ used)}
  \\
 \hline
 \ea
 \\
 \ \ \ \
 \stackrel{{\rm map\ } \Omega^{-1}}{}
 \ \
  \swarrow\ \  \
  \ \  \ \ \
  \ \ \ \ \ \ \
  \ \ \ \ \
  \ \ \ \ \ \ \
 \ \ \ \ \  \searrow \nwarrow\
 \stackrel{ \rm equivalence.}{}\\
 \begin{array}{|c|}
 \hline
 \vspace{-0.5cm}\\
  {\rm  \fbox{ {friendly\ representation\ space}}}\\
  \ {\cal K}= {\cal K}_{mathematical}
  \\
   {\rm \ (unphysical)}
  \\  \hline
 \ea
 \stackrel{\ {\rm simplification \ } \Theta\to I }{ \longleftarrow }
 \begin{array}{|c|}
 \hline
 \vspace{-0.5cm}\\
  {\rm  \fbox{ {ultimate\ amended\ picture}}}\\
  \ {\cal H}= {\cal H}_{physical}
  \\
 {\rm \ (represented\ in} \ {\cal K})
  \\
    \hline
 \ea
\\
\\
 \ea
 \label{ufin}
 \ee
In such an arrangement, by assumption, the
two lower-line Hilbert spaces ${\cal K}$
and  ${\cal H}$
coincide as linear modules
or vector-space sets
of the ket-vector elements
$|\psi\kt$.
Thus,
we can write $|\psi\kt \in {\cal K}$ and/or
$|\psi\kt \in {\cal H}$ and treat
the Hamiltonian $H$
(carrying the
input information about dynamics)
as an operator defined and
acting
in both of these two
spaces.

The difference
between ${\cal K}$
and  ${\cal H}$
lies in two conventions. First,
the former, auxiliary, manifestly
unphysical  Hilbert space ${\cal K}$
is definitely preferred as the user-friendlier one.
The standard Dirac's notation is applied to
the bra-vector elements of its dual marked by a prime,
$\br \psi| \in {\cal K}'$.
Second,
the correct physical Hilbert space ${\cal H}$ is only
treated as represented in ${\cal K}$ using
the mere change of the inner product,
 $$
 \br \psi_a|\psi_b\kt \ = {\rm the \ inner \ product\ in} \ {\cal K}\,,\ \ 
 \br \psi_a|\Theta|\psi_b\kt \ = {\rm the \ (mimicked) \  product\ in} \ {\cal H}\,.
 $$
The bra-vector elements of the dual physical
vector space are,
in the notation of Ref.~\cite{SIGMAdva},
denoted as ``brabras'',
$\bbr \psi| \in {\cal H}'$. They
have the metric-dependent
representation  $$\bbr \psi|= \br \psi|\Theta$$ in ${\cal K}'$.
Thus,
we can treat these brabras
as the Hermitian conjugates of the kets
in the physical
Hilbert space ${\cal H}$. In parallel, we can also
introduce
the
``ketkets'' $| \psi \kkt =\Theta\,|\psi\kt $
as the Hermitian conjugates of the brabras
with respect to the simpler, conventional inner products
in the unphysical but preferred
representation-Hilbert-space ${\cal K}$.

\section*{Appendix B. Rayleigh-Schr\"{o}dinger
construction in ${\cal L}$}

A factorization $\Theta=\Omega^\dagger\Omega$
of the metric enables us to define the
textbook ${\cal L}-$space self-adjoint avatar
of our
Hamiltonian
 \be
 \mathfrak{h}=\Omega\,H\,\Omega^{-1}=\mathfrak{h}^\dagger \,.
 \label{tarela}
 \ee
It acts in
the upper component ${\cal L}$
of diagram (\ref{ufin}) which
is just the conventional
physical Hilbert space of textbooks.
The latter Hilbert space 
can be perceived
as the set of the
``spiked-ket'' elements
 $
 |\psi\pkt = \Omega\,|\psi\kt \in {\cal L}
 $
and of their Hermitian-conjugate ``spiked-bra'' duals
 $
 \pbr \psi| = \br \psi|\, \Omega^\dagger \in {\cal L}'
 $.
By definition, the hypothetical and
practically inaccessible
operator $\mathfrak{h}$
is an ${\cal L}-$space image
of our preselected Hamiltonian $H$.
Hence,
the the above-mentioned links of
${\cal L}$ to the other two
spaces
imply that
the Hermiticity of
$\mathfrak{h}$ in ${\cal L}$
is equivalent to
the (hidden) Hermiticity of our $H$ in ${\cal H}$.
In contrast, the same operator $H$
is non-Hermitian in
the mathematical manipulation space ${\cal K}$.

In the context of perturbation theory
with $\mathfrak{h}=\mathfrak{h}(\lambda)=\mathfrak{h}_0+\lambda \mathfrak{v}$
in Schr\"{o}dinger equation
 \be
 \mathfrak{h}(\lambda)\,|\psi_n(\lambda)\pkt =
 E_n(\lambda)\,|\psi_n(\lambda)\pkt\,,
  \ \ \ \ \
 n = 0, 1, \ldots\,
 \label{selfadjo}
 \ee
the standard power-series ansatz
for energies (\ref{20})
is complemented by its wave-function analogue
 \be
 |\psi_n(\lambda)\pkt =|\psi_n(0)\pkt + \lambda\,|\psi_n^{(1)}\pkt
 +\lambda^2\,|\psi_n^{(2)}\pkt+\ldots \,.
 \label{20a}
 \ee
The Hermiticity
(\ref{tarela}) is then an important mathematical advantage.
In particular, this property enables us to
treat the unperturbed Schr\"{o}dinger equation
 \be
 (\mathfrak{h}_0-E_n(0))|\psi_n(0)\pkt=0\,
 \label{[6]}
  \ee
as a standard eigenvalue problem, preferably
solvable in closed form. Next,  we may
recall any textbook and
write down the first-order-approximation extension of Eq.~(\ref{[6]}),
 \be
 (\mathfrak{h}_0-E_n(0))|\psi_n^{(1)}\pkt+
 (\mathfrak{v}-E_n^{(1)})|\psi_n(0)\pkt
 =0\,
 \label{[7]}
  \ee
as well as its second-order extension
 \be
 (\mathfrak{h}_0-E_n(0))|\psi_n^{(2)}\pkt+
 (\mathfrak{v}-E_n^{(1)})|\psi_n^{(1)}\pkt+
 (-E_n^{(2)})|\psi_n(0)\pkt
 =0\,
 \label{[8]}
  \ee
etc.
In this manner we may reconstruct the
sequence of the corrections to the energy,
 \be
 E_n^{(1)}=\pbr \psi_n(0)|\mathfrak{v}|\psi_n(0)\pkt
 \,,
 \label{40a}
 \ee
 \be
 E_n^{(2)}=
 \pbr \psi_n(0)|\mathfrak{v}|\psi_n^{(1)}\pkt
 \,
 \label{40ab}
 \ee
(etc) as well as
the analogous sequence of the corrections
to the wave-function ket-vectors
 \be
 |\psi_n^{(1)}\pkt =
 Q\frac{1}{E_n(0)-Q\mathfrak{h}_0 Q}Q\mathfrak{v}|\psi_n(0)\pkt
 \,,
 \label{40aw}
 \ee
 \be
 |\psi_n^{(2)}\pkt =
 Q\frac{1}{E_n(0)-Q\mathfrak{h}_0 Q}Q(\mathfrak{v}-E_n^{(1)})|\psi_n^{(1)}\pkt
 \,
 \label{40abw}
 \ee
(etc)
where
the symbol
 \be
 Q= I -
 |\psi_n(0)\pkt \pbr \psi_n(0)|\,
 \ee
denotes an elementary projector ``out of model space''.



\section*{Appendix C. Open questions behind quasi-Hermitian perturbations
\label{sec3}}

In the ultimate physical Hilbert space ${\cal H}$
in which $H$ is self-adjoint
it would be possible to
introduce a dedicated superscript marking the space-characterizing
conjugation
and to rewrite Eq.~(\ref{zasdieu}) as follows,
 $$
 H = H^\ddagger :=\Theta^{-1}\,H^\dagger\,\Theta\,.
 $$
Nevertheless,
once we move to the preferred representation space ${\cal K}$
the latter notation becomes redundant
because the relation $H = H^\ddagger$
finds its rephrasing
in the
quasi-Hermiticity constraint~(\ref{zasdieu}) in
${\cal K}$.

In applications we have to
re-read Eq.~(\ref{zasdieu}) as
restricting
an assignment of
metric $\Theta$ to
a preselected non-Hermitian
operator $H$. Such a metric will necessarily vary with
the Hamiltonian in general, $\Theta=\Theta(H)$.
The same observation applies to
its Dyson-map factor, $\Omega=\Omega(H)$.
Both of these comments have already
been formulated in \cite{response}.
We pointed out there that
whenever one decides to consider any one-parametric family of
Hamiltonians $H=H(\lambda)$
[including also the perturbed Hamiltonians of Eq.~(\ref{perham})
as special case],
the physical meaning of the
quantum system
can only be deduced from its textbook
probabilistic interpretation in ${\cal L}$. At every $\lambda$.

This means that
the change of the parameter will
imply the change of diagram (\ref{ufin}).
The independence
of the unperturbed and perturbed versions of Schr\"{o}dinger
Eq.~(\ref{sekvoj})
leads to the necessity of working, at every non-vanishing
parameter $\lambda$,
with as many as six
separate Hilbert spaces.
Even though we can
merge ${\cal L}(\lambda)={\cal L}(0)={\cal L}$
and use the single textbook space for reference,
the union of the two respective
diagrams (\ref{ufin}) with $\lambda=0$ and $\lambda \neq 0$ still
has to be replaced by their five-Hilbert-space concatenation,
%
 \be
   \ba
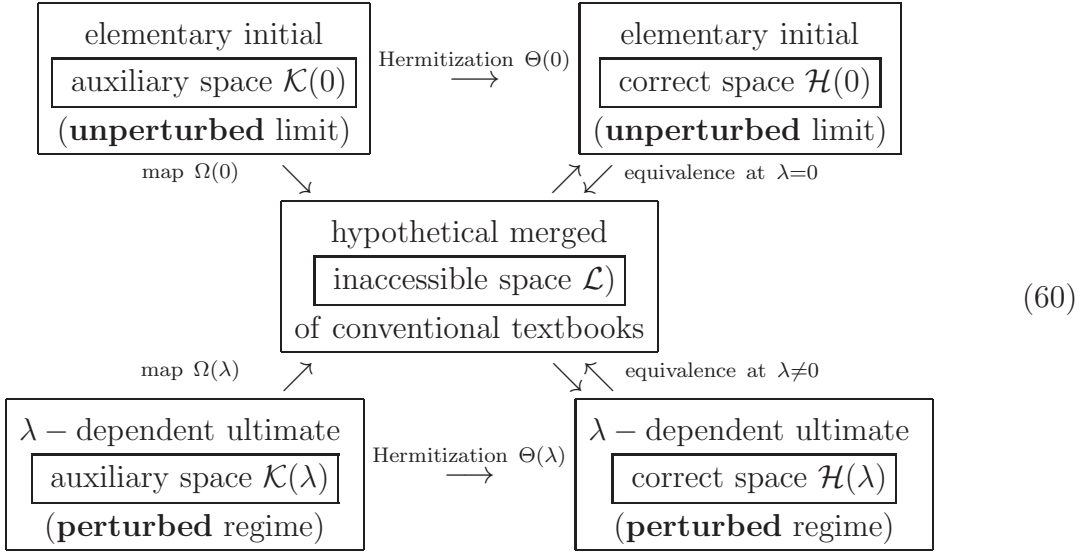
   \begin{array}{|c|}
 \hline
 \vspace{-0.5cm}\\  {\rm \ {elementary\ initial}}\
  \\
  {\rm  \fbox{ {auxiliary\ space} $ {\cal K}(0)$}}
  \\
  {\rm  {(\bf unperturbed\rm \ limit)}}
  \\
     \hline
 \ea
 \stackrel{ {\rm Hermitization \ } \Theta(0) }{ \longrightarrow }
 \begin{array}{|c|}
 \hline
 \vspace{-0.5cm}\\ {\rm \ {elementary\ initial}}\
  \\
  {\rm  \fbox{ {correct\ space} $ {\cal H}(0)$}}
  \\
  {\rm  {(\bf unperturbed\rm \ limit)}}
  \\
    \hline
 \ea
   \\
 \ \
 \stackrel{{\rm map\ } \Omega(0)}{}
 \ \ \ \
  \searrow\ \  \
  \ \  \
  \ \ \ \ \
  \ \ \ \ \ \ \
 \ \ \ \ \  \nearrow \swarrow\
 \stackrel{ {\rm equivalence\ at\ } \lambda=0}{}
 \\
     \begin{array}{|c|}
 \hline
 \vspace{-0.5cm}\\ {\rm \ {hypothetical}\ merged\ }
  \\
  {\rm  \fbox{ {inaccessible\ space} $ {\cal L})$}}
  \\
  {\rm  {of\ conventional\ textbooks}}
  \\
 \hline
 \ea
  \\
 \ \
 \stackrel{{\rm map\ } \Omega(\lambda)}{}
 \ \ \ \
  \nearrow\ \  \
  \ \  \
  \ \ \ \ \
  \ \ \ \ \ \ \
 \ \ \ \ \  \searrow \nwarrow\
 \stackrel{ {\rm equivalence\ at\ } \lambda \neq 0}{}
  \\
  \begin{array}{|c|}
 \hline
 \vspace{-0.5cm}\\ \lambda-{\rm dependent\ {ultimate}}\
  \\
  {\rm  \fbox{ {auxiliary\ space} $ {\cal K}(\lambda)$}}
  \\
  {\rm  {(\bf perturbed\rm \ regime)}}
  \\
     \hline
 \ea
 \stackrel{ {\rm Hermitization \ } \Theta(\lambda) }{ \longrightarrow }
 \begin{array}{|c|}
 \hline
 \vspace{-0.5cm}\\ {\rm \lambda-{\rm dependent\ {ultimate}}}\
  \\
  {\rm  \fbox{ {correct\ space} $ {\cal H}(\lambda)$}}
  \\ {\rm  {(\bf perturbed\rm \ regime)}}
  \\
    \hline
 \ea
 \\
 \\
 \ea
 \label{5ufin}
 \ee
In \cite{response}
we emphasized that
the general QHQM formalism
remains consistent and applicable even when
the $\lambda-$dependence of the Hilbert
space metric $\Theta(\lambda)$
is not smooth.
Nevertheless, we proposed that
for the perturbed models
of Eq.~(\ref{perham})
characterized by a smooth $\lambda-$dependence of the Hamiltonian
it makes sense to postulate also the
analyticity of $\Theta(\lambda)$.
Still,
our concluding comments concerning
the practical feasibility of the
calculations were skeptical.
In contrast,
one of the key messages as delivered by
our present paper
can be seen in a significant suppression of the latter
skepticism.


\section*{Appendix D. Biorthonormalized unperturbed bases}

In Hilbert space ${\cal K}$
our Hamiltonians
are assumed non-Hermitian, $H\neq H^\dagger$.
In section \ref{whava} we emphasized that
we have to complement, therefore, the conventional Schr\"{o}dinger
equation (i.e., Eq.~(\ref{sekvoj})
for the ket vectors)
by its conjugate partner specifying
their ${\cal H}-$space duals.
This goal is achieved either via Eq.~(\ref{Lnonselfadjo})
for the ``brabra'' vectors or, equivalently, via Eq.~(\ref{2nonselfadjo})
for the ``ketket'' vectors.

Temporarily let us simplify the mathematics and
assume that dim ${\cal K}< \infty$ \cite{Nimrodd}.
Then, for the reasons explained in diagram (\ref{5ufin}) of Appendix C
we must distinguish between the equations
at $\lambda=0$ (the unperturbed limit) and
at $\lambda\neq 0$ (the perturbed regime).
In the former case let us now
rewrite
both of the $\lambda=0$
Schr\"{o}dinger equations in a more compact notation,
 \be
 H\,|n\kt =E_n\,|n\kt\,,
 \ \ \ \ \
 n=0, 1, \ldots, \dim {\cal K}-1 \,,
 \label{prv}
 \ee
 \be
 H^\dagger\,|n\kkt =E_n\,|n\kkt\,,
 \ \ \ \ \
 n=0, 1, \ldots, \dim {\cal K}-1\,.
 \label{druh}
 \ee
In the framework of perturbation theory in its
most elementary form
the solutions of
such an advanced, ``doubled'' quasi-Hermitian bound-state problem
are usually assumed available
in closed form.
We will also require
that all of the unperturbed eigenvectors
form a biorthonormalized
set (i.e., one has $\bbr \psi_m|\psi_n\kt = \delta_{mn}$)
which is complete. Thus,
we will have, at our disposal,
the spectral decomposition of the identity operator,
 \be
 \sum_{n=0}^{\dim {\cal K}-1}\,|n\kt \,\bbr n| = I\,.
 \ee
Formally, one can even postulate the validity of
a spectral representation of the unperturbed Hamiltonian,
 \be
 H(0)=\sum_{n=0}^{\dim {\cal K}-1}\,|n\kt \,E_{n}(0)\,\bbr n|\,.
 \ee
Finally, recalling \cite{SIGMAdva} one can write down also
the multiparametric definition
 \be
 \Theta(0)=\sum_{n=0}^{\dim {\cal K}-1}\,|n\kkt \,|\kappa_{n}(0)|^2\,\bbr
 n|\,
 \label{13}
 \ee
of all of the metrics which would be formally compatible with
the Dieudonn\'{e}'s quasi-Hermiticity constraint (\ref{zasdieu})
at $\lambda=0$.
In parallel, the related Dyson-map factor $\Omega=\Omega(0)$
appearing in Eq.~(\ref{fakju}) and
in diagrams (\ref{ufin}) and/or (\ref{5ufin})
as well as in the
explicit definition $|\psi\pkt = \Omega\,|\psi\kt \in {\cal L}$
of the elements of the hypothetical space of textbooks
can be formally represented by the sum
 \be
 \Omega(0)=\sum_{n=0}^{\dim {\cal K}-1}\,|n\pkt \kappa_{n}(0)\,\bbr
 n|\,.
 \ee
Depending on the representation one can insert here
$\bbr \psi| \in {\cal H}'$
or  $\bbr \psi| = \br \psi|\Theta \in {\cal K}'$.


\newpage

\begin{thebibliography}{99}



\bibitem{BB}
Bender, C. M.; Boettcher, S.
Real Spectra in Non-Hermitian Hamiltonians Having
PT Symmetry.
\emph{Phys. Rev. Lett.} \textbf{1998}, \emph{80}, 5243.




 \bibitem{BG}
Buslaev, V.; Grecchi, V. {Equivalence of unstable anharmonic
oscillators and double wells.} \emph{J. Phys. A Math. Gen.}
\textbf{1993}, \emph{26}, 5541--5549.
%
%


\bibitem{Carl}
Bender, C. M. Making sense of non-Hermitian Hamiltonians. \emph{Rep.
Prog. Phys.} {\bf 2007}, {\em 70}, 947--1118.


\bibitem{Carlbook}
Bender, C. M.
(with contributions from
P. E. Dorey, C. Dunning, A. Fring,
D. W. Hook, H. F. Jones, S. Kuzhel, G. Levai, and R. Tateo).
 \emph{PT Symmetry in Quantum and Classical Physics};
World Scientific: Singapore, 2018.

\bibitem{Christodoulides}
Christodoulides, D.; Yang, J.-K. (Eds.)
\emph{Parity-Time Symmetry and Its Applications};
Springer: Singapore, 2018.

\bibitem{Nimrodc}
Bagchi, B.; Ghosh, R.; Sen, S.
Analogue Hawking Radiation as a Tunneling in a Two-Level
PT-Symmetric
System.
 \emph{Entropy}
 {\bf 2023},
 {\em 25},
      1202.


\bibitem{Stone}
Stone, M.H. On one-parameter unitary groups in Hilbert Space.
\emph{Ann. Math.} {\bf 1932}, {\em 33}, 643--648.


\bibitem{Geyer}
%
Scholtz, F. G.;  Geyer, H. B.;
Hahne,  F. J. W.
Quasi-Hermitian Operators in Quantum Mechanics and the Variational Principle.
\emph{Ann. Phys. (NY)} {\bf 1992}, {\em 213}, 74--101.



\bibitem{Dieudonne}
Dieudonne, J. Quasi-Hermitian Operators. In
\emph{Proc. Int. Symp. Lin. Spaces}, Pergamon: Oxford, UK, 1961,
{pp. 115--122}.
%


\bibitem{ali}
Mostafazadeh, A. Pseudo-Hermitian Representation of
Quantum Mechanics.
{\emph{Int. J. Geom. Meth. Mod. Phys.}} \textbf{2010}, \emph{7}, 1191--1306.



\bibitem{book}
%
Bagarello, F.; Gazeau, J.-P.; Szafraniec, F.; Znojil, M. (Eds.)
\emph{Non-Selfadjoint Operators in Quantum Physics: Mathematical Aspects};
Wiley: {Hoboken, NJ, USA,}
2015.



\bibitem{BM}
Bender, C. M.; Milton, K. A.
Nonperturbative Calculation of Symmetry Breaking
in Quantum Field Theory.
\emph{Phys. Rev.}
{\bf 1997},
{\em D 55},
       R3255.
%
%






\bibitem{Wu}
Bender, C. M.; Wu, T. T.
Anharmonic Oscillator.
\emph{Phys. Rev.} {\bf 1969}, {\em 184}, 1231 - 1260.


\bibitem{Turbiner}
Turbiner, A.; del Valle, D. C.
Anharmonic oscillator: a solution.
 \emph{J. Phys. A Math. Theor.}
\textbf{2021}, \emph{54}, 295404.


\bibitem{Kato}
Kato, T. \emph{Perturbation Theory for Linear Operators}; Springer:
Berlin/Heidelberg, Germany,
 1966.



\bibitem{response}
Znojil, M.
Theory of response to perturbations in non-hermitian systems
using five-Hilbert-space reformulation of unitary quantum mechanics.
 \emph{Entropy}
 {\bf 2020},
 {\em  22},
      80.

\bibitem{Trefethen}
Trefethen, L. N.; Embree, M.
\emph{Spectra and Pseudospectra: The Behavior of Nonnormal Matrices and Operators};
Princeton University Press: Princeton, 2005.



\bibitem{Brody}
Brody, D. C.
Biorthogonal quantum mechanics.
\emph{J. Phys. A: Math. Theor.}
{\bf 2013},
{\em 47},
      035305.



\bibitem{twoSE}
Znojil, M.
Quasi-Hermitian formulation of quantum mechanics
using two conjugate Schroedinger equations.
 \emph{Axioms}
 {\bf 2023},
 {\em  12},
      644.

%

\bibitem{Dyson}
Dyson, F.J. General theory of spin-wave interactions. \emph{ Phys.
Rev.} \textbf{1956}, \emph{102}, 1217.


\bibitem{Jensen}
Janssen, D.; D\"{o}nau, F.; Frauendorf, S.; Jolos, R. V. Boson
description of collective states. \emph {Nucl. Phys. A}
\textbf{1971}, \emph {172}, 145 - 165.


\bibitem{Mateo}
%
Jones, H. F.;  Mateo, J.
An Equivalent Hermitian Hamiltonian for the non-Hermitian $-x^4$
Potential.
\emph{Phys. Rev.} {\bf 2006}, {\em D 73},  085002.
%



\bibitem{Rafa}
Fern\'andez, F.; Guardiola, R.; Ros J.;
Znojil, M.
Strong-coupling expansions for the PT-symmetric oscillators
$V (r) = a i x +
b (ix)^2 + c (ix)^3$.
 \emph{J. Phys. A Math. Gen.}
\textbf{1998}, \emph{31},  10105 - 10112.



\bibitem{BD}
Bender, C. M.; Dunne,G, V,
Large-order perturbation theory
for a non-Hermiitan PT-symmetric Hamiltonian.
 \emph{J. Math. Phys.}
 {\bf 1999},
 {\em  40},
      4616 -- 4621.


\bibitem{Dysonb}
Koukoutsis, E.;
Hizanidis, K.;
Ram, A. K.;
Vahala, G.;
Dyson maps and unitary evolution for Maxwell equations in tensor dielectric media.
 \emph{Phys. Rev. A}
 {\bf 2023},
 {\em 107},
      042215.

\bibitem{Jones}
Jones, H. F. Interface between Hermitian and non-Hermitian Hamiltonians
in a model calculation.
%
%
\emph{Phys. Rev.} \textbf{2008}, \emph{D 78}, 065032.

\bibitem{Jonescomm}
Znojil, M. Scattering theory using smeared non-Hermitian
potentials.
%
\emph{Phys. Rev. D} \textbf{2009}, \emph{80},
045009.
%

\bibitem{Messiah}
Messiah, A. \emph{Quantum Mechanics};
North Holland: Amsterdam, The Netherlands, 1961.
%
%




\bibitem{RST}
Roch, S.; Silberman, B. $C^*$-algebra techniques
in numerical analysis.
%
 \emph{J. Operator Th.}
 {\bf 1996},
 {\em  35},
      241 -- 280.

\bibitem{Viola}
Krej\v{c}i\v{r}\'{\i}k, D.; Siegl, P.; Tater, M.; Viola, J.
Pseudospectra in non-Hermitian quantum mechanics.
\emph{J. Math. Phys. } \textbf{2015}, \emph{56}, 103513.


\bibitem{Horacek}
Rotter, I. {A non-Hermitian Hamilton operator and the physics
of open quantum systems}.
 \emph{J. Phys. A: Math. Theor. }
 {\bf 2009},
 {\em  42},
       153001.


\bibitem{Nimrod}
Moiseyev, N. \emph{Non-Hermitian Quantum Mechanics};
 Cambridge Univ. Press: Cambridge, 2011.


\bibitem{NIPa}
Jakubsk\'{y}, V.
Thermodynamics of Pseudo-Hermitian Systems in Equilibrium.
 \emph{Mod. Phys. Lett. A}
 {\bf 2007},
 {\em 22},
      1075 -- 1084.
%


\bibitem{NIP}
Znojil, M. Non-Hermitian interaction representation
and its use in relativistic quantum mechanics.
\emph{Ann. Phys. (NY)} {\bf 2017}, {\em 385}, 162--179.



\bibitem{NIPb}
Moise, A. A. A.;
Cox, G.;
Merkli, M.
Entropy and entanglement in a bipartite quasi-Hermitian system
and its Hermitian counterparts.
 \emph{Phys. Rev. A}
 {\bf 2023},
 {\em 108},
      012223.

\bibitem{Caliceti}
Caliceti, E.; Graffi, S.; Maioli, M.
 Perturbation theory of odd anharmonic oscillators.
 \emph{Commun.
Math. Phys.} \textbf{1980}, \emph{75}, 51 -- 66.


%


%
%


\bibitem{Simon}
Eremenko, A.; Gabrielov, A. {Analytic continuation of eigenvalues of
a quartic oscillator.} \emph{Comm. Math. Phys.} \textbf{2009},
\emph{287}, 431.


\bibitem{Liu}
Liu, Y. X.;
%
Jiang, X. P.;
%
Cao, J. P.;
%
Chen, S.
%
%
Non-Hermitian mobility edges in one-dimensional quasicrystals with
parity-time symmetry.
 \emph{Phys. Rev. B}
 {\bf 2020},
 {\em 101},
      174205.


\bibitem{SIGMA}
Znojil, M. Three-Hilbert-space formulation of Quantum Mechanics.
\emph{Symm. Integ. Geom. Meth. Appl. SIGMA} {\bf 2009}, {\em 5}, 001
%
(arXiv: 0901.0700).


\bibitem{SIGMAb}
Ju, C. Y.; Miranowicz, A.; Chen, Y. N.; Chen, G. Y.; Nori, F.
Emergent parallel transport and curvature in Hermitian and
non-Hermitian quantum mechanics.
 \emph{Quantum}
 {\bf 2024},
 {\em  8},
      1277.


\bibitem{CGbook}
Caliceti, E.; Graffi, S.
Criteria for the Reality of the Spectrum
of PT-Symmetric Schr\"{o}dudinger Operators.
In
Bagarello, F.; Gazeau, J.-P.; Szafraniec, F.; Znojil, M. (Eds.)
\emph{Non-Selfadjoint Operators in Quantum Physics: Mathematical Aspects};
Wiley: {Hoboken, NJ, USA,}
2015, ch. 4, pp. 189 - 240.



\bibitem{Nimroda}
Zezyulin,
D. A.;
Kartashov,
Y. V.;
Konotop, V. V.
Metastable two-component solitons near an exceptional
point.
  \emph{Phys. Rev. A}
 {\bf 2021},
 {\em 104},
      023504.



\bibitem{Nimrode}
Bagchi, B.; Ghosh, R.; Sen, S.
Exceptional point in a coupled Swanson system.
 \emph{EPL}
 {\bf 2022},
 {\em 137},
      50004.


\bibitem{EP3}
Guria, C.;
Zhong, Q.;
Ozdemir, S. K.;
Patil, Y. S. S.;
El-Ganainy, R.;
Harris, J. G. E.
Resolving the topology of encircling multiple exceptional points.
 \emph{Nature Communications}
 {\bf 2024},
 {\em  15},
       1369.

\bibitem{EPsa}
Henry, R. A.; Batchelor, M. T. Exceptional points in the
Baxter-Fendley free parafermion model.
%
%
 \emph{Scipost Phys.}
 {\bf 2023},
 {\em  15},
       016.

%


\bibitem{Siegl}
Siegl, P.; Krej\v{c}i\v{r}\'{\i}k, D.
On the metric operator for the imaginary cubic oscillator.
\emph{ Phys. Rev.} \textbf{2012}, \emph{D 86},  121702(R).
%
%
%
%


\bibitem{Makris}
Berry, M. V. Physics of Nonhermitian Degeneracies.
\emph{Czech. J. Phys.} \textbf{2004}, \emph{54}, 1039--1047.



\bibitem{Connes}
 Bagarello, F.; Fring, A.
 {A non selfadjoint model on a two dimensional noncommutative
 space
with unbound metric}.
  \emph{Phys. Rev. }
 {\bf 2013},
 {\em  A 88},
      042119.



\bibitem{SKbook}
Krej\v{c}i\v{r}\'{\i}k, D.; Siegl, P.
Elements of spectral theory without the spectral theorem.
In
Bagarello, F.; Gazeau, J.-P.; Szafraniec, F.; Znojil, M. (Eds.)
\emph{Non-Selfadjoint Operators in Quantum Physics: Mathematical Aspects};
Wiley: {Hoboken, NJ, USA,}
2015, ch. 5, pp. 241 - 292.

\bibitem{Uwe}
G\"{u}nther, U.; Stefani. F.
\emph{IR-truncated PT -symmetric $ix^3$ model and its asymptotic
spectral scaling graph}; arXiv 1901.08526,
\textbf{2019}.

\bibitem{denish}
{Ramirez, R.; Reboiro, M.; Tielas, D.}
Exceptional Points from the Hamiltonian of a hybrid physical system:
Squeezing and anti-Squeezing.
\emph{Eur. Phys. J. D} \textbf{2020}, \emph{74}, 193.



\bibitem{gride}
Brody, D. C.; Hughston, L. P.
Quantum measurement of space-time events.
\emph{J. Phys. A: Math. Theor.}
 {\bf 2021},
 {\em 54},
     235304.

\bibitem{ozky}
Alase, A.; Karuvade, S.; Scandolo, C. M.
The operational foundations
of PT-symmetric and quasi-Hermitian
quantum theory.
\emph{J. Phys. A: Math. Theor.} \textbf{2022}, \emph{55}, 244003.



\bibitem{Joshua}
Feinberg, J.; Riser, B.
Pseudo-Hermitian random-matrix models: General formalism.
\emph{Nucl. Phys.}
{\bf 2022},
{\em B 975},
      115678.


\bibitem{Semoradova}
Semor\'{a}dov\'{a}, I.; Siegl, P.
Diverging eigenvalues in domain truncations of Schroedinger
operators with complex potentials.
\emph{SIAM J. Math. Anal.} \textbf{2022}, \emph{54}, 5064-5101.


\bibitem{timedep}
Znojil, M.
Time-dependent version of cryptohermitian quantum theory.
\emph{ Phys. Rev.} \textbf{2008}, \emph{D 78}, 085003.


\bibitem{SIGMAdvab}
Wang, W. H.;
Chen, Z. L.;
Li, W,;
The metric operators for pseudo-Hermitian Hamiltonian.
 \emph{ANZIAM J.}
 {\bf 2023},
 {\em 65},
      215 -- 228.


\bibitem{SIGMAdva}
Znojil, M.
On the role of the normalization factors $\kappa_n$ and of
the pseudo-metric P in crypto-Hermitian quantum models.
\emph{Symm. Integ. Geom. Meth. Appl. SIGMA} {\bf 2008}, {\em 4}, 001
(arXiv: 0710.4432v3).


\bibitem{Nimrodd}
Ballesteros, A.;
Ram\'{\i}rez, R,;
Reboiro, M.
Non-standard quantum algebras and finite dimensional PT -symmetric
systems.
  \emph{J. Phys. A: Math. Theor.}
 {\bf 2024},
 {\em 57},
      035202.


\end{thebibliography}
\end{document}